\RequirePackage{fixltx2e}
\documentclass[aps,prb,reprint,floatfix]{revtex4-1}

\usepackage{amsmath,amsfonts,amssymb} %
\usepackage{mathtools} %
\usepackage{wasysym} %
\usepackage{bm} %
\usepackage[bbgreekl]{mathbbol} %
\usepackage{graphicx} %
\usepackage[dvipsnames,table]{xcolor} %
\usepackage{tabularx} %
\usepackage[acronym]{glossaries} %
\usepackage{braket} %
\usepackage[binary-units=true,detect-weight=true]{siunitx} %
\usepackage{fancyhdr} %
\usepackage{microtype} %
\usepackage{natmove}
\usepackage[byname]{smartref} %
\usepackage[breaklinks, %
colorlinks, linkcolor=Blue, citecolor=Blue, urlcolor=Blue]{hyperref} %
\usepackage[version=3]{mhchem}
\AtBeginDocument{\usepackage{booktabs}} 
\usepackage{tikz}
\usetikzlibrary{backgrounds,shadows.blur,fit,decorations.pathreplacing,shapes}

\DeclarePairedDelimiter{\floor}{\lfloor}{\rfloor} %
\newcommand{\E}{\textrm{e}} %
\newcommand{\I}{\mathrm{i}\mkern1mu} %
\newcommand{\RR}{\mathbf{R}} %
\newcommand{\rr}{\mathbf{r}} %
\newcommand{\eq}[1]{Eq.~(\ref{#1})} %
\def\be{\begin{equation}} %
\def\ee{\end{equation}} %
\def\bea{\begin{eqnarray}} %
\def\eea{\end{eqnarray}} %

\def\MyTitle{Qubit coupled-cluster method: A systematic approach to
  quantum chemistry on a quantum computer} %
\def\MyAuthora{Ilya G. Ryabinkin} %
\def\MyAuthorb{Tzu-Ching Yen} %
\def\MyAuthorc{Scott N. Genin} %
\def\MyAuthord{Artur F. Izmaylov} %
\newacronym{QPE}{QPE}{quantum phase estimation} %
\newacronym{VQE}{VQE}{variational quantum eigensolver} %
\newacronym{UCC}{UCC}{unitary coupled-cluster} %
\newacronym{QCC}{QCC}{qubit coupled-cluster} %
\newacronym{FCI}{FCI}{full configurational interaction} %
\newacronym{CASCI}{CASCI}{complete active space configurational
  interaction} %
\newacronym{JW}{JW}{Jordan--Wigner} %
\newacronym{BK}{BK}{Bravyi--Kitaev} %
\newacronym[longplural={degrees of freedom}, %
firstplural={degrees of freedom (DOF)}, plural={DOF}]{DOF}{DOF}{degree
  of freedom} %
\newacronym[longplural={equations of motion}, %
firstplural={equations of motion (EOM)}, %
plural={EOM}]{EOM}{EOM}{equation of motion} %
\newacronym{PES}{PES}{potential energy surface} %
\newacronym{CI}{CI}{configuration interaction} %
\newacronym{QMF}{QMF}{qubit mean-field} %
\newacronym{SQP}{SQP}{sequential quadratic programming} %
\newacronym{RHF}{RHF}{restricted Hartree--Fock}

\begin{document}

\title{\MyTitle}

\author{\MyAuthora{}} %
\affiliation{Department of Physical and Environmental Sciences,
  University of Toronto Scarborough, Toronto, Ontario\, M1C\,1A4,
  Canada}
\affiliation{Chemical Physics Theory Group, Department of Chemistry,
  University of Toronto, Toronto, Ontario\, M5S 3H6, Canada}

\author{\MyAuthorb{}} %
\affiliation{Department of Physical and Environmental Sciences,
  University of Toronto Scarborough, Toronto, Ontario\, M1C\,1A4,
  Canada}
\affiliation{Chemical Physics Theory Group, Department of Chemistry,
  University of Toronto, Toronto, Ontario\, M5S 3H6, Canada}

\author{\MyAuthorc{}} %
\affiliation{OTI Lumionics Inc., 100 College St. \#351, Toronto,
  Ontario\, M5G 1L5, Canada} %

\author{\MyAuthord{}} %
\affiliation{Department of Physical and Environmental Sciences,
  University of Toronto Scarborough, Toronto, Ontario\, M1C 1A4,
  Canada}
\affiliation{Chemical Physics Theory Group, Department of Chemistry,
  University of Toronto, Toronto, Ontario\, M5S 3H6, Canada}

\date{\today}

\begin{abstract}
  A \gls{UCC} form for the wavefunction in the variational quantum
  eigensolver has been suggested as a systematic way to go beyond the
  mean-field approximation and include electron correlation in solving
  quantum chemistry problems on a quantum computer. Although being
  exact in the limit of including all possible coupled-cluster
  excitations, practically, the accuracy of this approach depends on
  how many and what kind of terms are included in the wavefunction
  parametrization. Another difficulty of \gls{UCC} is a growth of the
  number of simultaneously entangled qubits even at the \emph{fixed}
  fermionic excitation rank. Not all quantum computing architectures
  can cope with this growth. To address both
  problems we introduce a \gls{QCC} method that starts directly in the
  qubit space and uses energy response estimates for ranking the
  importance of individual entanglers for the variational energy
  minimization. Also, we provide an exact factorization of a unitary
  rotation of more than two qubits to a product of two-qubit unitary
  rotations. Thus, the \gls{QCC} method with the factorization
  technique can be limited to only two-qubit entanglement gates and
  allows for very efficient use of quantum resources in terms of the
  number of coupled-cluster operators. The method performance is
  illustrated by calculating ground-state potential energy curves of
  H$_2$ and \ce{LiH} molecules with chemical accuracy, $\le 1$\,
  kcal/mol.
\end{abstract}

\glsresetall

\maketitle

\section{Introduction}

A many-body fermionic problem defining electronic properties of 
molecules and materials from first-principles can be written as 
the following parametric time-independent Schr\"odinger equation
\begin{equation}
  \label{eq:el_se}
  \hat H_e(\RR) \ket{\Psi_i(\RR)} = E_i(\RR) \ket{\Psi_i(\RR)}.
\end{equation}
Here, $\RR=(\RR_1,...\RR_{N_n})$ are nuclear variables treated as parameters,
$E_i(\RR)$ and $\ket{\Psi_i(\RR)}$ are \glspl{PES} and adiabatic electronic states, respectively,
and $\hat H_e(\RR)$ is the electronic Hamiltonian of the system 
\begin{eqnarray}
  \notag
  \hat H_e(\RR) & = & -\frac{1}{2}\sum_{i=1}^{N_e} \nabla_{\rr_i}^2
                      + \sum_{i<j}^{N_e}\frac{1}{|\rr_i-\rr_j|} \\
                & + & \sum_{\alpha<\beta}^{N_n}
                      \frac{1}{|\RR_\alpha-\RR_\beta|} - \sum_{i,\alpha}\frac{1}{|\rr_i-\RR_\beta|},
\end{eqnarray}
where $\rr_i$ are the electronic variables, $N_e$ and $N_n$ are number
of electrons and nuclei.
Computational cost of the exact numerical solution of \eq{eq:el_se}
scales exponentially with the size (e.g. the number of electrons) of
the system on a classical computer. To address this issue without
introducing approximations, it was proposed to employ a quantum
computer~\cite{Lloyd:1996/sci/1073,AspuruGuzik:2005/sci/1704}, which
may solve the problem with polynomial scaling of computational cost
with the system size.

However, immediate application of quantum computers is hampered by
the limited quantity and quality of available quantum resources. In
particular, contemporary architectures offer only few ($<10$) fully interacting qubits, 
and their quantum coherent state can exist for limited time. As can be
gauged both from theoretical and experimental efforts, scaling-up the
quantum computer is far from straightforward. Even anticipating
noticeable growth of capabilities of quantum hardware in near future,
to become competitive with solving the electronic structure problem on classical 
computers, it is crucial to develop computationally frugal algorithms utilizing 
present and near future quantum computers in full.

Several algorithmic developments in this direction have already taken
place. In order to formulate the electronic structure problem for a
quantum computer that operates with qubits (two-level systems), the
electronic Hamiltonian needs to be transformed iso-spectrally to its
qubit form. This is done in two steps. First, the
second quantized form of $\hat H_e$ is obtained
\begin{equation}
  \label{eq:qe_ham}
  \hat H_e = \sum_{pq} h_{pq}(\RR) {\hat a}^\dagger_p {\hat a}_q + \frac{1}{2}\sum_{pqrs}
  g_{pqrs}(\RR) {\hat a}^\dagger_p {\hat a}^\dagger_q {\hat a}_s {\hat a}_r,
\end{equation}
where ${\hat a}_p^\dagger$ (${\hat a}_p$) are fermionic creation
(annihilation) operators, $h_{pq}(\RR)$ and $g_{pqrs}(\RR)$ are one-
and two-electron integrals in a spin-orbital
  basis.\cite{Helgaker:2000} This step has polynomial complexity and
is carried out on a classical computer. Then, using the
\gls{JW}~\cite{Jordan:1928/zphys/631, AspuruGuzik:2005/sci/1704} or
more resource-efficient \gls{BK}
transformation~\cite{Bravyi:2002/aph/210, Seeley:2012/jcp/224109,
  Tranter:2015/ijqc/1431, Setia:2017/ArXiv/1712.00446,
  Havlicek:2017/pra/032332}, the electronic Hamiltonian is converted
iso-spectrally to a qubit form
\begin{equation}
  \label{eq:spin_ham}
  \hat H = \sum_I C_I(\RR)\,\hat P_I,
\end{equation}
where $C_I(\RR)$ are numerical coefficients, and $\hat P_I$ are
Pauli  ``words", products of Pauli operators of different qubits 
\begin{equation}
  \label{eq:Pi}
  \hat P_I = \cdots \hat w_{1}^{(I)} \, \hat w_{0}^{(I)},
\end{equation}
$\hat w_i^{(I)}$ is one of the $\hat x,\hat y,\hat z$ Pauli
operators for the $i^{\rm th}$ qubit.

The \gls{QPE} algorithm~\cite{Abrams:1997/prl/2586,
  Abrams:1999/prl/5162, AspuruGuzik:2005/sci/1704} was historically
the first approach that relied on estimating the phase of the
stationary-state wavefunction using the quantum Fourier
transform.\cite{Nielsen:2010} For good accuracy one would need to
perform a long coherent evolution of the system for accurate phase
estimation. Practical difficulties in maintaining qubits' coherence
for long time hinder the use of the \gls{QPE} approach for quantum
chemistry simulations on currently available quantum computers.

This prompted a development of a more resource-efficient method, the
\gls{VQE}~\cite{Peruzzo:2014/ncomm/4213, Wecker:2015/pra/042303},
where the trial ground-state \emph{stationary} wavefunction $\Psi$ is
prepared by an action of a suitably chosen unitary operator $\hat U$
on an initial state of qubits $\ket{\Phi}$,
$\ket{\Psi} = \hat U \ket{\Phi}$. $\hat U$ is optimized variationally
to provide an estimate for the ground state exact energy
\begin{equation}
  \label{eq:vqe}
  E_0 \le \braket{\hat U \Phi| \hat H | \hat U\Phi}. 
\end{equation}
Of course, there is a great deal of freedom in parametrization of
$\hat U$. The original proposal of
Ref.~\citenum{Peruzzo:2014/ncomm/4213} was to borrow the
parametrization of $\hat U$ from the fermion-to-qubit
  transformation of a fermionic \gls{UCC} ansatz,
\begin{equation}
  \label{eq:ucc_param}
  \hat U_{\rm CC} = \exp(\hat T - {\hat T}^\dagger),
\end{equation}
where
\begin{align}
  \label{eq:T_def}
  \hat T = & \hat T_1 + \hat T_2 + \ldots + \hat T_{N_e}, \\
  \label{eq:T1_def}
  \hat T_1 = & \sum_{ia} t_i^a \hat a_a^\dagger \hat a_i, \\
  \label{eq:T2_def}
  \hat T_2 = & \sum_{ijab} t_{ij}^{ab} \hat a_a^\dagger \hat a_b^\dagger \hat a_j
               \hat a_i, \\
           & \ldots \nonumber
\end{align}
Here $\hat T_k$ are mutually commuting fermionic $k$-fold (single,
double, \emph{etc.}) excitation operators that promote electrons from
occupied orbitals $i, j, \ldots$ to unoccupied (virtual) orbitals
$a, b, \ldots$ in a reference single-determinant $\ket{\Phi}$ and
parametrized by the corresponding amplitudes
$t_i^a,\ t_{ij}^{ab}, \ldots$.
$\hat T$ may be truncated at rank $m \le N_e$, leading to an approximate
treatment, but at any rank greater than 1 the \gls{UCC} ansatz is
exponentially hard for a classical computer due to non-truncation of
the Baker--Campbell--Hausdorff series\cite{Taube:2006/ijqc/3393}. On
the other hand, it is known that coupled cluster-type expansions are
quickly convergent, especially if electron correlation is not too
strong\cite{Olsen:2000/jcp/7140}, thus, there is a hope to find
accurate energy estimates already at the low rank.

Yet, even though superiority of the \gls{UCC}-based \gls{VQE} approach
over \gls{QPE} has been convincingly
demonstrated\cite{OMalley:2016/prx/031007}, the computational cost of
\gls{UCC} still grows relatively quickly even at low orders. The main
difficulty is that all fermion-to-qubit transformations increase the
length of Pauli words involved in $\hat U$. In other words, even if
only a one-body cluster operator, $\hat T_1$, is used for
$\hat U_{\rm CC}$, its qubit representation is a unitary operator
involving Pauli words of the length $\sim \log(N_o)$ (in the case of
the \gls{BK} transformation, others increase the length even faster),
where $N_o$ is the number of one-electron spin-orbitals in the second
quantized form of $\hat H_e$. For some lattice models there are
special transformations that limits this growth of
\emph{locality}\cite{Havlicek:2017/pra/032332}, but for a general
molecular Hamiltonian the best count is still logarithmic. This
becomes a challenge because multi-qubit operations require either
entangling of increasing number of qubits (not all quantum computers
can do that) or factorizing the multi-qubit operation as a product of
one- and two-qubit ones. The latter was shown to be possible in
principle, but there are no general approaches to construct an
efficient factorization proposed to date.\cite{Nielsen:2010}
A typical approximate approach involves Trotterization.\cite{Whitfield:2011/mp/735}
Frequently, fermionic excitation operators, Eqs.~(\ref{eq:T1_def}--\ref{eq:T2_def}), are transformed to a 
qubit form and Trotterized.\cite{Romero:2018/qct/preprint-1, Barkoutsos:2018/pra/022322}

Another strategy for constructing the wavefunction has been adopted recently in
Ref.~\citenum{Kandala:2017/nature/242}. The wavefunction ansatz was
tailored to hardware requirements and had the form
\begin{equation}
  \label{eq:ibm_vqe_ansatz}
  \begin{split}
    \Psi(\boldsymbol \theta) = &\prod_{q = 1}^{N_q}
    \left[U^{q,d}(\boldsymbol \theta)\right] \times U_\text{ENT}
    \times \cdots \\
    & \times \prod_{q = 1}^{N_q}\left[ U^{q,1}(\boldsymbol \theta)\right]
    \times U_\text{ENT} \times \prod_{q = 1}^{N_q}\left[
      U^{q,0}(\boldsymbol \theta)\right] \ket{00 \ldots 0},
  \end{split}
\end{equation}
where $N_q$ is the number of qubits, 
$\ket{00 \ldots 0}$ is an initial state of the qubits and is exact equivalent of 
$\ket{\Phi}$ in \eq{eq:vqe}.  
Equation~\eqref{eq:ibm_vqe_ansatz} is an alternating sequence of products
of individual qubit rotations $U^{q,j}(\boldsymbol \theta)$,
\begin{equation}
  \label{eq:U_def}
  U^{q,j}(\boldsymbol \theta) = \E^{i z_q\theta_1^{q, j}} \E^{i
    x_q\theta_2^{q, j}} \E^{i z_q\theta_3^{q, j}},\ 0 \le j \le d
\end{equation}
where $\theta_1^{q,j}$, $\theta_2^{q,j}$, and $\theta_3^{q,j}$ are the Euler
angles of the $q^{\rm th}$ qubit at the $j^{\rm th}$ level, interleaved with action of ``entanglers''
\begin{equation}
  \label{eq:entangler}
  U_\text{ENT}(\tau) = \exp(-i \tau \hat P),
\end{equation}
where $\tau$ is an amplitude, and $\hat P$ is a two-qubit Pauli word.
The number $d$ was called a depth of the scheme. The first set of
$z$-rotations in $U^{q,0}(\boldsymbol \theta)$ was not implemented
($\theta_3^{q, 0} = 0$, $q = 1 \ldots {N_q}$), and entanglers' amplitudes
$\tau$ were all kept fixed, so that the
ansatz~\eqref{eq:ibm_vqe_ansatz} contained ${N_q}(3d+2)$ variational
parameters in total. By minimizing the expectation value of the
Hamiltonian, Eq.~\eqref{eq:vqe}, with respect to these parameters for
sufficiently high depths, reasonably accurate ground-state electronic
energies as functions of internuclear distances for molecules H$_2$,
LiH, and BeH$_2$ were obtained.

The obvious advantage of the
  form~\eqref{eq:ibm_vqe_ansatz} is its minimal degree of entanglement, only two qubits at a time. 
  The unitary transformation is given directly
in the qubit space, in contrast to the \gls{UCC} form that starts with 
fermionic operators. Perhaps, this fact alone is responsible
for considerable reduction of quantum resources needed for simulations
and eventually allowed for successful simulations of molecules larger
than \ce{H_2}. The rigorous basis for convergence of
ansatz~\eqref{eq:ibm_vqe_ansatz} to the exact answer with increasing
depth $d$ is the theorem stating that any multi-qubit unitary
transformation can be presented as a product of one- and two-qubit
unitary transformations.\cite{Polak:2011} However, 
  efficient construction of such a factorization is yet unsolved
  problem. Therefore, it is clear that the
form~\eqref{eq:ibm_vqe_ansatz} can be improved if some specifics of
the problem is taken into account. First, it can be shown that only
two individual qubit rotation angles per each qubit are sufficient at
any depth $d$, provided that entanglers' amplitudes are fully
optimized. Second, a set of pre-defined generators of entanglement
$\hat P$ [Eq.~\eqref{eq:entangler}] may not lead to the fastest
convergence---it is the system which defines the most ``efficient''
entanglers, not the hardware.

In the current paper we introduce a new systematic approach, a \gls{QCC} method, 
within the \gls{VQE} formalism. \gls{QCC} resembles a
coupled-cluster hierarchy, but contrary to \gls{UCC}, it is built directly
in the qubit space. Therefore, we bypass any explicit fermionic
constructions and save as much quantum resources as possible. To
achieve systematic convergence, we introduce an entanglers'
screening protocol that ranks various
  entanglers according to their estimated contributions into the
  correlation energy. To allow for an arbitrary multi-qubit 
  entanglers on limited-qubit hardware, we derive an exact
  factorization formula that represents a multi-qubit entangler as 
 a product of two-qubit entanglers. This
factorization uses commutation relations of the qubit Lie algebra and
introduce the minimal number of terms. We
illustrate its essential features on ground potential energy surfaces
of the H$_2$ and \ce{LiH} molecules, which are small
enough to be amenable for a comprehensive study, while at the same time 
are complex enough to illustrate many quantum-chemical
phenomena, including the transition to the strong-correlation limit
upon dissociation.

\section{Theory}
\label{sec:theory}

\subsection{Qubit Coupled-Cluster method}
\label{sec:wave-funct-param}

The \gls{QCC} wavefunction has the form
\begin{equation}
  \label{eq:qcc_gen}
  \Psi(\boldsymbol \tau, \boldsymbol \Omega) = \hat U(\boldsymbol
  \tau) \ket {\boldsymbol \Omega},
\end{equation}
where $\ket {\boldsymbol \Omega}$ and $\hat U (\boldsymbol \tau)$
represent the mean-field and correlation parts of a wavefunction.
The mean-field wavefunction is a product of single-qubit coherent
states\cite{Radcliffe:1971/jpa/313, Arecchi:1972/pra/2211,
  Perelomov:1972, Lieb:1973/cmp/327} 
\bea
  \label{eq:mf_wf}
  \ket {\boldsymbol \Omega} &=& \prod_{i=1}^{N_q} \ket {\Omega_i},\\%
  \label{eq:spin_coh_state_qubit}
  \ket{\Omega_i} &=& \cos\left(\frac{\theta_i}{2}\right) \ket{\alpha}
  + \E^{\I\phi_i}\sin\left(\frac{\theta_i}{2}\right) \ket{\beta}, \eea
  where $\phi_i$ and $\theta_i$ are azimuthal and polar angles on the
  ``Bloch sphere'' of the $i^{th}$ qubit, respectively, and
  $\ket{\alpha}$ and $\ket{\beta}$ are spin-up and spin-down
  eigenstates of the $\hat s_z(i) = \hat z_i/2$ operator. The
  single-qubit coherent states constitute a normalized but
  non-orthogonal complete set.
Correlation is introduced by multi-qubit rotations parametrized by 
real-valued amplitudes $\boldsymbol \tau = \{\tau_k\}$ as
\begin{equation}
  \label{eq:U_qcc_def}
  \hat U(\boldsymbol \tau) = \prod_{k=1}^{N_{\rm ent}} \exp(-i \tau_k \hat P_k/2),
\end{equation}
where $\hat P_k$ are the Pauli words [\eq{eq:Pi}] whose lengths vary
from 2 to $N_q$, $N_{\rm ent}$ is less or equal to the total number of
possible $\hat P_k$ operators, $4^{N_q} - 3N_q-1$, and a factor of
$1/2$ is introduced for convenience.

The expectation value of the Hamiltonian for the \gls{QCC}
parametrization [\eq{eq:qcc_gen}] is
\begin{align}
  \label{eq:barE}
  E(\boldsymbol \tau, \boldsymbol \Omega) & = \braket{\Psi(\boldsymbol
                                            \tau, \boldsymbol \Omega)|\hat
                                            H|\Psi(\boldsymbol \tau,
                                            \boldsymbol \Omega)}  \\
  \label{eq:barE_alt}
                                          & = \braket{\boldsymbol
                                            \Omega| U(\boldsymbol \tau)^\dagger \hat H U(\boldsymbol \tau) | \boldsymbol \Omega},
\end{align}
and minimization of $E(\boldsymbol \tau, \boldsymbol \Omega)$ with respect to 
amplitudes $\boldsymbol \tau$ and angles in $\boldsymbol \Omega$ provides the ground state 
energy. The main difference between \gls{QCC} and \gls{UCC} forms is in the
nature of generators of unitary transformations. In the \gls{UCC}
formalism generators are \emph{sums} of mutually-commuting fermionic
operators [Eqs~(\ref{eq:T_def}--\ref{eq:T2_def})], each of those, in
turn, after the \gls{JW} or \gls{BK} transformation is a
lengthy linear combination of Pauli words. This leads to excessive
consumption of precious quantum resources, for example, two-particle
entangling gates. Moreover, exponent of such quantities are not known
in closed form and, thus, are not amenable for extensive analysis. In contrast, 
the $\hat P_k$ generators of the \gls{QCC} ansatz are simple Pauli
words, and thus, all of them are involutory operators, $\hat P_k^2 = 1$. 
This property provides a closed form for  
similarity transformed Hamiltonians of individual entanglers 
\begin{align}
  \label{eq:qcc_func_single_tau}
  \hat H[\tau;\hat P] & = \E^{\I \tau \hat P/2}\hat H \E^{-\I \tau \hat P/2} \nonumber \\
          & = \hat H - \I \frac{\sin \tau}{2} [\hat H, \hat P] +
            \frac{1}{2}\left(1 - \cos\tau\right) \hat P\, [\hat H, \hat P].
\end{align}
Hence, it is obvious that the fully transformed Hamiltonian 
$\tilde H = U(\boldsymbol \tau)^\dagger \hat H U(\boldsymbol \tau)$
in \eq{eq:barE_alt} will involve $\sim 3^{N_{\rm ent}}$ operators thus revealing the exponential
complexity of the \gls{QCC} form. 

Overall, the \gls{QCC} scheme requires
$2N_q + N_\text{ent}$ variational parameters, where $2N_q$ is the number
of Bloch angles for $N_q$ qubits, and $N_\text{ent}$ is the number of
optimized entanglers' amplitudes.

One possible concern about using Pauli words as generators is potential breaking of
the fermionic symmetry by a trial electronic function and ensuing
unphysical behavior of energy an properties. Indeed, an approximated
wavefunction is prone to such a phenomenon\cite{Lowdin:1963/rmp/496},
and as we found in Ref.~\citenum{Ryabinkin:2018/unpub-cnstr}, it is
more a rule than an exception. However, the \gls{QCC} formalism can be
equipped with constraints, such as the particle number or electronic spin, in a
straightforward manner to guarantee conservation of proper physical
symmetries.

\subsection{Entanglers' ranking}
\label{sec:pert-select-entangl}

The key quantity in our ranking of all, $4^{N_q}-3N_q-1$ entanglers for the $N_q$-qubit
 system is the similarity-transformed energy function,
\begin{equation}
  \label{eq:ent_ranking_criterion}
E[\tau; \hat P_k] = \min_{\boldsymbol \Omega} \braket{\boldsymbol
    \Omega | e^{i\tau\hat P_k/2}\hat H e^{-i\tau \hat P_k/2} | \boldsymbol \Omega}.
\end{equation}
 Using the closed form of the similarity transformed
Hamiltonian, \eq{eq:qcc_func_single_tau}, it is straightforward to
evaluate $E[\tau; \hat P_k]$ and to find the energy lowering due to
variation of $\tau$
\begin{align}
  \label{eq:ent_energy_lowering}
  \Delta E[\hat P_k] & = \min_{\tau} E[\tau; \hat P_k] - E[0; \hat
                       P_k] \nonumber \\
                     &  = \min_{\tau} E[\tau; \hat P_k] - E_\text{QMF}
                       \le 0.
\end{align}
Here $E_\text{QMF} = \min_{\boldsymbol \Omega} \braket{\boldsymbol
  \Omega | \hat H | \boldsymbol \Omega}$ is the \gls{QMF} energy, which is  
  equivalent to $E[0; \hat P_k]$. 
  Thus, all entanglers can be ranked according to their $\Delta E[\hat P_k]$ values. 
  The main downside of this procedure is its computational cost: it requires full
optimization of both $\tau$ and the mean-field angles $\boldsymbol \Omega$ for 
each entangler that can be expensive for large systems.

To alleviate this problem, a pre-screening based on the first and
second terms of the Taylor expansion of individual similarity-transformed energies
[Eq.~\eqref{eq:ent_ranking_criterion}] 
\begin{align}
  \label{eq:12der}
  E[\tau; \hat P_k] =  E_\text{QMF} & + \tau \frac{\mathrm{d}E[\tau;
                      \hat P_k]}{\mathrm{d}\tau}\Big|_{\tau = 0}
                                      \nonumber \\
                                    & + \frac{\tau^2}{2}
                                      \frac{\mathrm{d}^2E[\tau; \hat
                                      P_k]}{\mathrm{d}\tau^2}\Big|_{\tau
                                      = 0} + \ldots
\end{align}
can be done. One caveat in calculating these derivatives is accounting for the
relaxation of the mean-field part due to changes in $\tau$.
Essentially, for every value of $\tau$ the minimizing set of Bloch
angles is different, therefore they are implicit functions $\tau$.
Thus, the full $\tau$ derivative in \eq{eq:12der} can be expanded as
\begin{equation}
  \label{eq:full_tau_der}
  \frac{\mathrm{d}}{\mathrm{d}\tau} = \frac{\partial}{\partial \tau} +
  \sum_{i = 1}^{2N_q} \left(\frac{\mathrm{d} \varphi_i} {\mathrm{d} \tau}\right)_\text{min}
  \frac{\partial}{\partial \varphi_i},
\end{equation}
where $\{\varphi_i\}_{i=1}^{2N_q}$ denotes $\{\theta_1, \phi_1,...\theta_{N_q}, \phi_{N_q},\}$
the set of the Bloch angles parametrizing the mean-field solution. It is straightforward to show (see the Appendix)
that the first derivative at $\tau=0$ is given by:
\begin{equation}
  \label{eq:PT1_test}
  \frac{ \mathrm{d} E[\tau;\hat P_k]}{\mathrm{d} \tau} \Big|_{\tau = 0} =
  \Braket{{\boldsymbol \Omega}_\text{min} | -\frac{\I}{2}[\hat H, \hat P_k] |
    {\boldsymbol \Omega}_\text{min}},
\end{equation}
where $\ket{\boldsymbol \Omega_\text{min}}$ is a coherent state
evaluated on the set of optimized at the mean-field level angles
$\{\varphi_i\}$. Evaluation of the second order derivative in
\eq{eq:12der} is more involved and is detailed in the Appendix.

Thus, the important entanglers can be ranked within two tiers. The
first tier is formed by the entanglers with non-zero absolute values
of their first energy derivatives [\eq{eq:PT1_test}]. The second tier
includes the entanglers with vanishing first energy derivatives but
significant {\it negative} second energy derivatives,
Eq.~\eqref{eq:2nd_der_explicit}. The final ranking is done based on
evaluating $\Delta E[\hat P_k]$ values for top entanglers in both tiers.

\subsection{Factorization of multi-qubit entanglers}
\label{sec:fact-multi-qubit}

Existing quantum hardware is typically limited in the form of
operators it can efficiently implement. One of such limitations is the
maximum number of qubits that are possible to entangle by realizable
unitary transformations, in other words, the length of the Pauli word
[\eq{eq:Pi}] generating a unitary transformation is usually limited to
two qubits. 
Thus, if our ranking procedure identifies three- or four-body
entanglers as contributing significantly to the energy lowering, it
may not be possible to directly implement them due to hardware
limitations.

Here, we show how to factorize a unitary transformation entangling
more than two qubits in a product of unitary transformations involving
only two-qubit generators. The factorization is a recursive procedure
applied at each step to a many-qubit unitary to produce three new
unitaries, each of them parametrized by fewer-qubits Pauli words.

The elementary factorization step can be illustrated as follows:
\begin{enumerate}
\item Assume we have Pauli word $\hat P$ with the length
  $|\hat P| \ge 3$, where the length $|\cdot|$ is defined as the
  number of Pauli operators in $\hat P$. Then $\hat P$ can be
  represented as
  \begin{equation}
    \label{eq:longPrep}
    \hat P = \hat P_1 \hat w_k \hat P_2,
  \end{equation}
  where the subscript of $\hat w_k$ denotes that this elementary Pauli
  operator corresponds to the $k^{\rm th}$ qubit. The word
  $\hat P_1$ ($\hat P_2$) contains all qubit indices that are strictly
  greater (lower) than $k$. $\hat P_1$, $\hat P_2$, and $\hat w_k$ are
  all mutually commutative, since they act on non-overlapping sets of
  qubits. Also by a choice of $k$ one can always satisfy the following
  relations
  \begin{equation}
    \label{eq:P12_leng}
    |\hat P_{1,2}| = \left\{
      \begin{array}{lc}
        \frac{|\hat P|-1}{2},
        & \quad |\hat P|\ \text{is odd} \\[1.5ex]
        \frac{|\hat P|}{2}\ \text{and}\ \frac{|\hat P|}{2}-1,
        & \quad |\hat P|\ \text{is even} 
      \end{array}
    \right. 
  \end{equation}

\item Let us denote $\hat w'_k$ and $\hat w''_k$ the Pauli operators
  for the $k^{\rm th}$ qubit that satisfy the following commutation
  relation with $\hat w_k$
  \begin{equation}
    \label{eq:w_commut}
    [\hat w'_k, \hat w''_k] = 2\I\, \hat w_k,
  \end{equation}
  which is always possible by properties of a single-qubit Pauli
  operators. Substituting $\hat w_k$ by its commutator in \eq{eq:longPrep}
  gives
  \begin{equation}
    \label{eq:P_fact}
    \hat P = -\frac{\I}{2} \left[\hat P_1 \hat w'_k, \hat w''_k \hat P_2\right].
  \end{equation}
  On the other hand, the same commutator as in \eq{eq:P_fact} can be
  obtained if the expression
  $\exp[{\I (\pi/4)\, \hat w''_k \hat P_2}] \hat P_1 \hat w'_k
  \exp[{-\I (\pi/4)\, \hat w''_k \hat P_2}]$ is considered using
  Eq.~\eqref{eq:qcc_func_single_tau}, and therefore
  \begin{equation}
    \label{eq:entangler_exp_transf}
    \hat P = \E^{\I (\pi/4)\, \hat w''_k \hat P_2} \hat P_1 \hat w'_k \E^{-\I (\pi/4)\, \hat w''_k \hat P_2}.
  \end{equation}
  
\item Finally, from Eq.~\eqref{eq:entangler_exp_transf} it follows
  that \bea \hat P^n = \E^{\I (\pi/4)\, \hat w''_k \hat P_2} (\hat P_1
  \hat w'_k)^n \E^{-\I (\pi/4)\, \hat w''_k \hat P_2}, \eea and
  therefore, exponentiation of $\hat P$ using the Taylor series gives
  the final factorization
\begin{equation}
    \label{eq:expP}
    \E^{-\I t \hat P} = \E^{\I (\pi/4)\, \hat w''_k \hat P_2} \E^{-\I t \hat P_1
      \hat w'_k} \E^{-\I (\pi/4)\, \hat w''_k \hat P_2}.
  \end{equation}
\end{enumerate}
Eq.~\eqref{eq:expP} expresses the unitary transformation with the
generator $\hat P$ as a product of three unitary transformations with
generators of at most $\floor{|\hat P|/2} + 1$ length. Thus, the
number of two-qubit factors grows as $\sim \log_2 |\hat P|$. 
Also, note that the factorization does not increase the number of variational parameters,
only the number of entanglers.

\section{Numerical studies and discussion}
\label{sec:numer-stud-disc}

We illustrate our developments by computing the ground state potential
energy curves within the \gls{QMF} and \gls{QCC} approaches for the
\ce{H2} and \ce{LiH} molecules. These molecules were used to
illustrate performance of quantum computing techniques
previously.\cite{Kandala:2017/nature/242,Hempel:2018/prx/031022}.
\ce{LiH} is particular interesting because it is one of the simplest
molecules, where electronic correlation entangles electrons on more
than two orbitals.

\subsection{Fermionic Hamiltonian quantities}
\label{sec:electr-struct-calc}

To generate fermionic Hamiltonian spin-orbitals and one- and two-electron integrals, 
a locally modified \textsc{gamess}\cite{gamessus-2} electronic structure
package was used.

Molecules were oriented along the $z$ axis, and canonical \gls{RHF}
orbitals were computed in the STO-3G atomic basis set\cite{EMSL-2}
assuming $D_{2h}$ and $C_{2v}$ symmetry for \ce{H2} and \ce{LiH},
respectively.\footnote{The full molecular symmetry group is
  $D_{\infty h}$ and $C_{\infty v}$, but the maximal Abelian subgroups
  with all-real irreducible representations were chosen instead for
  the sake of implementation simplicity.} For the \ce{H2} molecule all
4 spin-orbitals were taken into account in construction of the
second-quantized electronic Hamiltonian, Eq.~\eqref{eq:qe_ham}. For
the \ce{LiH} molecule we determine the subset of ``active''
spin-orbitals to produce the electronic Hamiltonian as follows. Out of
6 molecular orbitals of \ce{LiH}, four belong to $A_1$ and two
remaining correspond to the $B_1$ and $B_2$ irreducible
representations of the $C_{2v}$ group. The lowest-energy $A_1$-type
orbital and both $B$-type orbitals were frozen, which means their
populations were fixed at 2, 0, and 0, respectively. The remaining 3
orbitals of $A_1$ type constituted the active space where the
electronic Hamiltonian~\eqref{eq:qe_ham} was computed. It must be
noted that one of $A_1$ orbitals lies \emph{higher} in energy than
both $B$-type ones, so that reordering of the canonical \gls{RHF}
orbitals, which are typically sorted by orbital energy, was necessary.
Molecular integral transformation was carried out by the
\textsc{aldet} module; 
this module was also used to generate reference \gls{CASCI} energies.

\subsection{Generation of qubit operators}
\label{sec:gener-qubit-hamilt}

\paragraph*{H$_2$ molecule:}
One- and two-electron integrals in the canonical \gls{RHF} molecular
orbitals basis for each value of the internuclear distance, $R$(H-H), 
were used in the \gls{BK} transformation to produce the corresponding qubit Hamiltonians.
Spin-orbitals were alternating in the order $\alpha$, $\beta$,
$\alpha$, .... Although qubits 2 and 4 are known to be stationary, in
other words, all operators in the qubit Hamiltonian act by only $z$ or
the identity operators on these qubits, we did not use this additional
symmetry to decrease the qubit count in the qubit Hamiltonians, so
that all 4 qubits corresponding to 4 spin-orbitals were retained. The
number of electrons, $\hat N$, and the square of the total spin,
$\hat S^2$, operators were prepared from their second quantized
fermionic expressions.\cite{Helgaker:2000} These operators are
extremely useful in analysis of both \gls{QMF} and \gls{QCC}
solutions.
 
\paragraph*{LiH molecule:}
Qubit Hamiltonians for various $R$(Li-H) distances 
were generated using the parity
fermion-to-qubit transformation\cite{Nielsen:2005/scholar_text}.
Spin-orbitals were arranged as ``first all alpha then all beta'' in
the fermionic form; since there are 3 active molecular orbitals in the
problem, this lead to a 6-qubit Hamiltonian containing 118 Pauli words
for each $R$(Li-H). These qubit Hamiltonians have 3$^{rd}$ and 6$^{th}$
stationary qubits, which allows one to replace the corresponding
$\hat z$ operators by their eigenvalues, $\pm 1$, thus defining the
different ``sectors'' of the original Hamiltonian. Each of these
sectors is characterized by its own 4-qubit effective Hamiltonian. The
ground state lies in the $z_{2} = -1$, $z_{5} = 1$ (qubits are
enumerated from 0) sector; the corresponding effective Hamiltonian has
100 Pauli terms. Analogous reduction procedure has been applied to the
qubit images of $\hat N$ and $\hat S^2$ operators.

\subsection{Potential energy curves}
\label{sec:qmf-qcc-results}

\begin{figure}[b]
  \centering %
  \includegraphics[width=1.0\columnwidth]{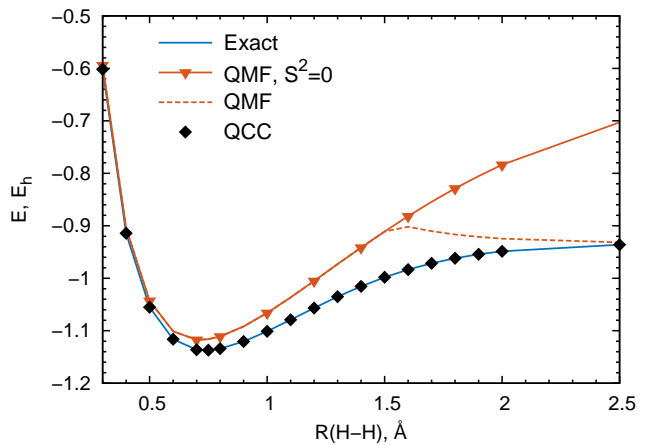}
  \caption{Potential energy curves for the H$_2$ molecule.}
  \label{fig:h2_curve}
\end{figure}

\begin{figure}
  \centering %
  \includegraphics[width=1.0\columnwidth]{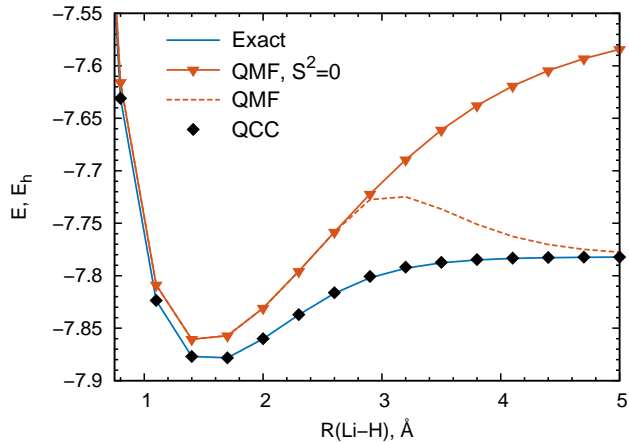}
  \caption{Potential energy curves for the LiH molecule.}
  \label{fig:lih_curve}
\end{figure}

As we established in Ref.~\citenum{Ryabinkin:2018/unpub-cnstr},
variational methods in qubit space are prone to symmetry breaking. This 
is particularly dramatically seeing in the \gls{QMF} case 
(Figs.~\ref{fig:h2_curve} and \ref{fig:lih_curve}). To restore the correct 
singlet spin symmetry and to make \glspl{PES} smooth, it is 
enough to imposed the spin constraint ($\braket{S^2} = 0$) on 
the \gls{QMF} solutions. Alternatively, more correlation can be added 
to \gls{QMF} to approach the exact solution, which has the correct singlet symmetry.    
In all our \gls{QCC} calculations we started with symmetry unconstraint 
\gls{QMF} solutions.

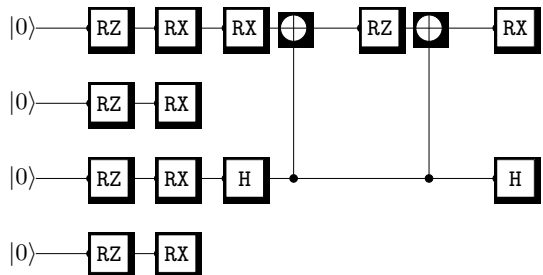
\begin{figure}
  \centering
  \begin{tikzpicture}[scale=1.0, transform shape]
\tikzstyle{basicshadow}=[blur shadow={shadow blur steps=8, shadow xshift=0.7pt, shadow yshift=-0.7pt, shadow scale=1.02}]\tikzstyle{basic}=[draw,fill=white,basicshadow]
\tikzstyle{operator}=[basic,minimum size=1.5em]
\tikzstyle{phase}=[fill=black,shape=circle,minimum size=0.1cm,inner sep=0pt,outer sep=0pt,draw=black]
\tikzstyle{none}=[inner sep=0pt,outer sep=-.5pt,minimum height=0.5cm+1pt]
\tikzstyle{measure}=[operator,inner sep=0pt,minimum height=0.5cm, minimum width=0.75cm]
\tikzstyle{xstyle}=[circle,basic,minimum height=0.35cm,minimum width=0.35cm,inner sep=0pt,very thin]
\tikzset{
shadowed/.style={preaction={transform canvas={shift={(0.5pt,-0.5pt)}}, draw=gray, opacity=0.4}},
}
\tikzstyle{swapstyle}=[inner sep=-1pt, outer sep=-1pt, minimum width=0pt]
\tikzstyle{edgestyle}=[very thin]
\node[none] (line0_gate0) at (0.2,-0) {$\Ket{0}$};
\node[phase] (line0_gate1) at (1.0999999999999999,-0) {};
\node[none] (line0_gate1) at (1.0999999999999999,-0) {};
\node[none,minimum height=0.5cm,outer sep=0] (line0_gate2) at (1.3499999999999999,-0) {};
\node[none] (line0_gate3) at (1.5999999999999999,-0) {};
\draw[operator,edgestyle,outer sep=0.5cm] ([yshift=0.25cm]line0_gate1) rectangle ([yshift=-0.25cm]line0_gate3) node[pos=.5]{\verb|RZ|};
\draw (line0_gate0) edge[edgestyle] (line0_gate1);
\node[phase] (line0_gate4) at (1.9999999999999998,-0) {};
\node[none] (line0_gate4) at (1.9999999999999998,-0) {};
\node[none,minimum height=0.5cm,outer sep=0] (line0_gate5) at (2.25,-0) {};
\node[none] (line0_gate6) at (2.5,-0) {};
\draw[operator,edgestyle,outer sep=0.5cm] ([yshift=0.25cm]line0_gate4) rectangle ([yshift=-0.25cm]line0_gate6) node[pos=.5]{\verb|RX|};
\draw (line0_gate3) edge[edgestyle] (line0_gate4);
\node[phase] (line0_gate7) at (2.9,-0) {};
\node[none] (line0_gate7) at (2.9,-0) {};
\node[none,minimum height=0.5cm,outer sep=0] (line0_gate8) at (3.15,-0) {};
\node[none] (line0_gate9) at (3.4,-0) {};
\draw[operator,edgestyle,outer sep=0.5cm] ([yshift=0.25cm]line0_gate7) rectangle ([yshift=-0.25cm]line0_gate9) node[pos=.5]{\verb|RX|};
\draw (line0_gate6) edge[edgestyle] (line0_gate7);
\node[none] (line1_gate0) at (0.2,-1) {$\Ket{0}$};
\node[phase] (line1_gate1) at (1.0999999999999999,-1) {};
\node[none] (line1_gate1) at (1.0999999999999999,-1) {};
\node[none,minimum height=0.5cm,outer sep=0] (line1_gate2) at (1.3499999999999999,-1) {};
\node[none] (line1_gate3) at (1.5999999999999999,-1) {};
\draw[operator,edgestyle,outer sep=0.5cm] ([yshift=0.25cm]line1_gate1) rectangle ([yshift=-0.25cm]line1_gate3) node[pos=.5]{\verb|RZ|};
\draw (line1_gate0) edge[edgestyle] (line1_gate1);
\node[phase] (line1_gate4) at (1.9999999999999998,-1) {};
\node[none] (line1_gate4) at (1.9999999999999998,-1) {};
\node[none,minimum height=0.5cm,outer sep=0] (line1_gate5) at (2.25,-1) {};
\node[none] (line1_gate6) at (2.5,-1) {};
\draw[operator,edgestyle,outer sep=0.5cm] ([yshift=0.25cm]line1_gate4) rectangle ([yshift=-0.25cm]line1_gate6) node[pos=.5]{\verb|RX|};
\draw (line1_gate3) edge[edgestyle] (line1_gate4);
\node[none] (line2_gate0) at (0.2,-2) {$\Ket{0}$};
\node[phase] (line2_gate1) at (1.0999999999999999,-2) {};
\node[none] (line2_gate1) at (1.0999999999999999,-2) {};
\node[none,minimum height=0.5cm,outer sep=0] (line2_gate2) at (1.3499999999999999,-2) {};
\node[none] (line2_gate3) at (1.5999999999999999,-2) {};
\draw[operator,edgestyle,outer sep=0.5cm] ([yshift=0.25cm]line2_gate1) rectangle ([yshift=-0.25cm]line2_gate3) node[pos=.5]{\verb|RZ|};
\draw (line2_gate0) edge[edgestyle] (line2_gate1);
\node[phase] (line2_gate4) at (1.9999999999999998,-2) {};
\node[none] (line2_gate4) at (1.9999999999999998,-2) {};
\node[none,minimum height=0.5cm,outer sep=0] (line2_gate5) at (2.25,-2) {};
\node[none] (line2_gate6) at (2.5,-2) {};
\draw[operator,edgestyle,outer sep=0.5cm] ([yshift=0.25cm]line2_gate4) rectangle ([yshift=-0.25cm]line2_gate6) node[pos=.5]{\verb|RX|};
\draw (line2_gate3) edge[edgestyle] (line2_gate4);
\node[phase] (line2_gate7) at (2.9,-2) {};
\node[none] (line2_gate7) at (2.9,-2) {};
\node[none,minimum height=0.5cm,outer sep=0] (line2_gate8) at (3.15,-2) {};
\node[none] (line2_gate9) at (3.4,-2) {};
\draw[operator,edgestyle,outer sep=0.5cm] ([yshift=0.25cm]line2_gate7) rectangle ([yshift=-0.25cm]line2_gate9) node[pos=.5]{\verb|H|};
\draw (line2_gate6) edge[edgestyle] (line2_gate7);
\node[xstyle] (line0_gate10) at (3.8000000000000003,-0) {};
\draw[edgestyle] (line0_gate10.north)--(line0_gate10.south);
\draw[edgestyle] (line0_gate10.west)--(line0_gate10.east);
\node[phase] (line2_gate10) at (3.8000000000000003,-2) {};
\draw (line2_gate10) edge[edgestyle] (line0_gate10);
\draw (line0_gate9) edge[edgestyle] (line0_gate10);
\draw (line2_gate9) edge[edgestyle] (line2_gate10);
\node[phase] (line0_gate11) at (4.7,-0) {};
\node[none] (line0_gate11) at (4.7,-0) {};
\node[none,minimum height=0.5cm,outer sep=0] (line0_gate12) at (4.95,-0) {};
\node[none] (line0_gate13) at (5.2,-0) {};
\draw[operator,edgestyle,outer sep=0.5cm] ([yshift=0.25cm]line0_gate11) rectangle ([yshift=-0.25cm]line0_gate13) node[pos=.5]{\verb|RZ|};
\draw (line0_gate10) edge[edgestyle] (line0_gate11);
\node[xstyle] (line0_gate14) at (5.6000000000000005,-0) {};
\draw[edgestyle] (line0_gate14.north)--(line0_gate14.south);
\draw[edgestyle] (line0_gate14.west)--(line0_gate14.east);
\node[phase] (line2_gate11) at (5.6000000000000005,-2) {};
\draw (line2_gate11) edge[edgestyle] (line0_gate14);
\draw (line0_gate13) edge[edgestyle] (line0_gate14);
\draw (line2_gate10) edge[edgestyle] (line2_gate11);
\node[phase] (line0_gate15) at (6.500000000000001,-0) {};
\node[none] (line0_gate15) at (6.500000000000001,-0) {};
\node[none,minimum height=0.5cm,outer sep=0] (line0_gate16) at (6.750000000000001,-0) {};
\node[none] (line0_gate17) at (7.000000000000001,-0) {};
\draw[operator,edgestyle,outer sep=0.5cm] ([yshift=0.25cm]line0_gate15) rectangle ([yshift=-0.25cm]line0_gate17) node[pos=.5]{\verb|RX|};
\draw (line0_gate14) edge[edgestyle] (line0_gate15);
\node[phase] (line2_gate12) at (6.500000000000001,-2) {};
\node[none] (line2_gate12) at (6.500000000000001,-2) {};
\node[none,minimum height=0.5cm,outer sep=0] (line2_gate13) at (6.750000000000001,-2) {};
\node[none] (line2_gate14) at (7.000000000000001,-2) {};
\draw[operator,edgestyle,outer sep=0.5cm] ([yshift=0.25cm]line2_gate12) rectangle ([yshift=-0.25cm]line2_gate14) node[pos=.5]{\verb|H|};
\draw (line2_gate11) edge[edgestyle] (line2_gate12);
\node[none] (line3_gate0) at (0.2,-3) {$\Ket{0}$};
\node[phase] (line3_gate1) at (1.0999999999999999,-3) {};
\node[none] (line3_gate1) at (1.0999999999999999,-3) {};
\node[none,minimum height=0.5cm,outer sep=0] (line3_gate2) at (1.3499999999999999,-3) {};
\node[none] (line3_gate3) at (1.5999999999999999,-3) {};
\draw[operator,edgestyle,outer sep=0.5cm] ([yshift=0.25cm]line3_gate1) rectangle ([yshift=-0.25cm]line3_gate3) node[pos=.5]{\verb|RZ|};
\draw (line3_gate0) edge[edgestyle] (line3_gate1);
\node[phase] (line3_gate4) at (1.9999999999999998,-3) {};
\node[none] (line3_gate4) at (1.9999999999999998,-3) {};
\node[none,minimum height=0.5cm,outer sep=0] (line3_gate5) at (2.25,-3) {};
\node[none] (line3_gate6) at (2.5,-3) {};
\draw[operator,edgestyle,outer sep=0.5cm] ([yshift=0.25cm]line3_gate4) rectangle ([yshift=-0.25cm]line3_gate6) node[pos=.5]{\verb|RX|};
\draw (line3_gate3) edge[edgestyle] (line3_gate4);
\end{tikzpicture}
\caption{A sample quantum circuit representing the \gls{QCC} ansatz
  for the \ce{H2} molecule on the Rigetti machine.} 
  \label{fig:circut-H2}
\end{figure}

\paragraph*{Entanglers in H$_2$:} We rank entanglers at
$R$(H-H) $= 1.0$\,\AA\ according to the procedure described in
Sec.~\ref{sec:pert-select-entangl} (the full table can be found in
Supplementary materials). Since only two out of four qubits contribute
nontrivially to the eigenstates of the H$_2$ qubit Hamiltonian,
two-qubit entanglers are sufficient for obtaining the exact solution. 
Out of total 54 two-qubit entanglers only 6 provide
energy lowering; all of them were found by the combined first-
[Eq.~\eqref{eq:PT1_test}] and second-order
[Eq.~\eqref{eq:2nd_der_explicit}] screening, see
Table~\ref{tab:h2_ranking}. Only 2 of them has non-zero energy
gradient and, thus, are identified by the cheapest first-order test,
Eq.~\eqref{eq:PT1_test}.
\begin{table}
  \centering
  \caption{Two-qubit entangler ranking for H$_2$ at $R = 1.0$\,\AA\ by
    Eqs.~(\ref{eq:PT1_test}), \eqref{eq:2nd_der_explicit}, and
    \eqref{eq:ent_energy_lowering}. Hartree atomic units.}
  \sisetup{%
    table-format = 3.4, %
    round-mode = places, %
    round-precision = 4} %
  \begin{tabularx}{1.0\linewidth}{@{\hfill}XSSS@{}}
    \toprule
    Entangler $\hat P$
    & {$\dfrac{\mathrm{d}E(\tau; \hat P)}{\mathrm{d}\tau}\Big|_{\tau = 0}$}
    & {$\dfrac{\mathrm{d^2}E(\tau; \hat P)}{\mathrm{d}\tau^2}\Big|_{\tau = 0}$}
    & {$\Delta E[P]$} \\ 
    \midrule
    ${\hat x}_2{\hat y}_0$ & -0.196791     &  0.5342956 & -0.035041681 \\
    ${\hat y}_2{\hat x}_0$ &  0.196791     &  0.5342956    & -0.035041681 \\
     ${\hat z}_2{\hat y}_0$ &  0.000000     & -0.0509  & -0.035041681 \\
     ${\hat z}_2{\hat x}_0$ &  0.000000     & -0.0509  & -0.035041681 \\
     ${\hat y}_2{\hat z}_0$ &  0.000000     &  -0.049314664 & -0.035041681 \\
     ${\hat x}_2{\hat z}_0$ &  0.000000     & -0.049314664  & -0.035041681 \\
    \bottomrule
  \end{tabularx}
  \label{tab:h2_ranking}
\end{table}
Moreover, since all 6 energy-lowering entanglers are related by a
simple global frame rotation (change of quantization axis) it is
possible to use any one of them 
for the whole range of $R(\ce{H-H})$, see
Fig.~\ref{fig:h2_curve}.
Thus, the \ce{H2} molecule in the minimal basis requires only one entangler to
reach the exact energy. This entangler is readily identified by our
ranking procedure without any \emph{ad hoc} assumptions about the
structure of a wave function or reference to fermionic operators.
A quantum computer circuit suitable for the Rigetti
machine~\cite{Rigetti:2017/1712.05771} that solves the \ce{H2} problem
by the \gls{QCC} method with the chosen entangler is shown in
Fig.~\ref{fig:circut-H2}. Besides a set of one-qubit gates, such as  
rotations around $x$ (\texttt{RX}) and $z$ (\texttt{RZ}) axis, and the Hadamard (\texttt{H}) gate,
it contains only two two-qubit \texttt{CNOT} gates. They
appear as a result of a decomposition (made by
\textsc{PyQuil}\cite{Pyquil:2016/1608.03355}) of the unitary
transformation generated by the entangler ${\hat x}_2{\hat y}_0$ into
elementary operations possible on the Rigetti machine. The two
leftmost rows of single-qubit gates encode the spin coherent states
and depend on the Bloch angles $(\phi_i, \theta_i),\ i = 1,\ldots, 4$,
the \texttt{RZ} gate interlaid between the two \texttt{CNOT} gates
depends on the entangler's amplitude $\tau$.

\begin{table}
  \centering
  \caption{Top entanglers for LiH at $R = 3.2$\,\AA\ by
    Eqs.~(\ref{eq:PT1_test}), \eqref{eq:2nd_der_explicit}, and
    \eqref{eq:ent_energy_lowering}. Hartree atomic units. $\Delta E[P]$'s are evaluated 
    with respect to the symmetry constrained QMF energies.}
  \sisetup{%
    table-format = 4.4, %
    round-mode = places, %
    round-precision = 4} %
  \begin{tabularx}{1.0\linewidth}{@{\hfill}XSS@{}}
    \toprule
    Entangler $\hat P$
    & {$\dfrac{\mathrm{d}E_{\rm cQMF}(\tau; \hat P)}{\mathrm{d}\tau}\Big|_{\tau = 0}$}
    & {$\Delta E[P]$} \\ 
    \midrule
    ${\hat x}_2{\hat x}_1{\hat y}_0$            & -0.079216013  & -0.090884078  \\
    ${\hat z}_3{\hat y}_2{\hat z}_1{\hat x}_0$  & -0.078106108  & -0.090861713 \\
    ${\hat x}_2{\hat y}_1{\hat x}_0$            & -0.079216012  & -0.090701597 \\
    ${\hat z}_3{\hat x}_2{\hat x}_1{\hat y}_0$  & -0.079216013  & -0.0895746   \\
    ${\hat x}_3{\hat y}_2{\hat z}_1{\hat x}_0$  & -0.079216012  & -0.0895746   \\
    ${\hat z}_3{\hat x}_2{\hat y}_1{\hat x}_0$  & -0.079216012  & -0.088714096 \\
    ${\hat y}_3{\hat x}_2{\hat z}_1{\hat x}_0$  & -0.079216012  & -0.088714096 \\
    \bottomrule
  \end{tabularx}
  \label{tab:lih_ranking}
\end{table} 

\paragraph*{Entanglers in LiH:} Due to a larger number of entangled
spin-orbitals in LiH, the list of entanglers is more complex. Near the
equilibrium bond length, $R(\ce{Li-H}) \approx 1.5$\,\AA, our
procedure readily identifies a few (6-7) entanglers that give
chemical accuracy. However, these entanglers do not provide chemical
accuracy for the whole curve, $R(\ce{Li-H})$ from 0.5 to 5.0\,\AA. The
region where QMF experiences symmetry breaking,
$R(\ce{Li-H}) \gtrapprox 3.2$\,\AA, is especially demanding: the list
of energetically-important entanglers grows substantially and
calculating the same entanglers' characteristics as before does not
shrink it considerably. To achieve reduction, we noticed that since
the QMF reference has broken symmetry while the exact eigenstate has
the singlet symmetry, it is prudent to search for entanglers that can
not only lower the energy but also restore the spin symmetry. To
assess the capability of entanglers for symmetry restoration we
calculate the gradient of individual energies on a spin-singlet
\emph{constraint} QMF solution,
$\ket{{\boldsymbol \Omega}_\text{c,min}}$:
\begin{equation}
  \label{eq:PT1_testC}
  \frac{ \mathrm{d} E_{\rm cQMF}[\tau;\hat P_k]}{\mathrm{d} \tau}
  \Big|_{\tau = 0} = \Braket{{\boldsymbol \Omega}_\text{c,min} |
    -\frac{\I}{2}[\hat H, \hat P_k] | {\boldsymbol
      \Omega}_\text{c,min}}.
\end{equation}
These gradients are larger for those entanglers that can affect the
symmetry of the constrained QMF solution. Using 7 of such
entanglers, which also have high individual energy lowerings
(Table~\ref{tab:lih_ranking}), we were able to reach the chemical
accuracy for the whole LiH potential energy curve
(Fig.~\ref{fig:lih_curve}). Interestingly, these entanglers that
 can not only restore symmetry but break it, work well for
nuclear geometries where symmetry restoration is not needed. We
attribute this to ability of the variational procedure to find
amplitudes for the symmetry affecting entanglers so that they do not
break QMF symmetry where it is unnecessary.

Applying our factorization procedure to split three- and four-qubit entanglers 
in Table~\ref{tab:lih_ranking} gives $31$  two-qubit entanglers. In
general, the decomposition overhead is logarithmic, but the exact
figures depend on the set of entanglers. The factorization does not change 
the number of variational parameters and their total number 
in the \gls{QCC} procedure for \ce{LiH} is only $2\times 4 + 7 = 15$.

If we compare our \gls{QCC} ansatz with alternatives from the
literature for the \ce{LiH} application, we find the following. If one
uses the \gls{QPE} method, order of 100 exponential factors---the size
of the Hamiltonian---are needed.

The simplified \gls{UCC} ansatz of
Ref.~\citenum{Hempel:2018/prx/031022} started from analysis of
fermionic double excitation operators [Eq.~\eqref{eq:T2_def}] inferred
from the configuration-interaction single and double excitation (CISD)
calculations. It was found 16 Pauli words are necessary to represent
only \emph{two} most important ones; to be able to proceed the authors
kept rather arbitrarily chosen two two-qubit entanglers out of this
list. 

The form of \citet{Kandala:2017/nature/242} seems to be the most
economical one: chemical accuracy was achieved with $d = 6$, 
which implies at least 6 pairwise entanglers.
However, \citet{Kandala:2017/nature/242} use significantly more
variational parameters: for 4 qubits and $d = 6$ the count is 80. The
excessive parametrization makes energy minimization considerably more
difficult since the expectation values provided by a quantum computer
have inherent quantum uncertainty.

\section{Conclusions}
\label{sec:conclusions}

We have introduced and studied a new method for performing electronic
structure calculations on a quantum computer. We called our method the
\acrfull{QCC} to emphasize that it is rooted in the \acrfull{UCC}
theory, but contrary to \gls{UCC}, formulated without any reference to
fermionic algebra exclusively in terms of qubit operators.

From the mathematical point of view, \gls{QCC} is a variational method
where variational optimization of single-qubit rotations and
multi-qubit unitary transformation are done altogether. The
single-qubit Hilbert space is parametrized in terms of spin coherent
states which depend on Bloch angles. There are only two angles per
qubit, and terms of the qubit Hamiltonian have exceptionally simple
expressions for matrix elements in the chosen basis.

The essential feature of the \gls{QCC} form is the multi-qubit unitary
transformations in terms of Pauli words---products of Pauli operators.
Such transformations are the simplest possible in the qubit tensor
algebra; they lead to a compact economical form of
unitarily-transformed operators like the Hamiltonian
itself~[Eq.~\eqref{eq:qcc_func_single_tau}]. This simplicity allows
for extremely frugal use of quantum resources such as two-qubit
entanglement gates. Moreover, we derived the decomposition of a
multi-qubit unitary rotation to a product of two-qubit unitary
rotations, Eq.~\eqref{eq:expP}. This formula can make simulations of
highly entangled systems possible on hardware that is limited by
entanglement of only two qubits.

Formulations of the quantum chemistry in the qubit space have several complications. 
First, even though only 
\emph{iso-spectral} fermionic-to-qubit transformations are employed,
and thus all symmetries of the fermionic system are replicated in its qubit
counterpart, this is not generally true for {\it approximate} methods.
Indeed, in Ref.~\citenum{Ryabinkin:2018/unpub-cnstr} we found that breaking of
physical symmetries, such as the number of electrons or spin,
in qubit variational methods is more a rule than an exception. At the
same time, this problem can be solved by using appropriate
constraints. Second, by treating a molecular system directly
in the qubit space, it is not straightforward to use typical electronic
structure intuition, for example, information about whether this system is
weakly-correlated like a closed-shell molecule near equilibrium
geometry, or strongly-correlated like an open-shell radical with many
quasi-degenerate orbitals. We address this shortcoming by devising the
derivative test, Eq.~\eqref{eq:PT1_test}, which allows one to tailor
the \gls{QCC} ansatz to a given molecular system in an efficient
manner.

We illustrate our approach by considering the
dissociation of the H$_2$ and \ce{LiH} molecules in the minimal basis set. 
\ce{LiH} contains more orbitals than H$_2$, and hence, more qubits are needed, 
but what is more important, the quantum entanglement in LiH is considerably more complex
than in H$_2$.
Even with this complexity we identified only seven generators of
entanglement, Pauli words, that are needed for the \gls{QCC}
parametrization, Eq.~\eqref{eq:qcc_gen}, to achieve chemical accuracy 
for the entire ground state curve. 
This results in only 15 variational parameters---8 Bloch angles and 7
generators' amplitudes. These numbers are considerably lower, in many
cases by few orders of magnitude, than the number of similar
quantities used by other methods. We hope, therefore, that the
\gls{QCC} method will become the method of choice for quantum
chemistry on a quantum computer in near future.

\acknowledgments %
A.F.I. acknowledges financial support from Natural Sciences and
Engineering Research Council of Canada (NSERC) through the Engage Plus
grant.

\appendix*
\section{Qubit coupled-cluster energy derivatives}
\label{sec:qubit-coupl-clust}

Let us define $\widetilde H(t) = \E^{-\I t \hat P/2} \hat H \E^{-\I t\hat P/2}$,
then the \gls{QCC} energy function
Eq.~\eqref{eq:ent_ranking_criterion} can be compactly written as
\begin{equation}
  \label{eq:ent_ranking_formula_appendix}
  E(t) = \min_{\boldsymbol \Omega} \Braket{\boldsymbol \Omega |
    \widetilde H(t) | \boldsymbol \Omega}. 
\end{equation}
The minimization condition makes the Bloch angles functions of $t$. It
is convenient to introduce $\ket{\boldsymbol \Omega_\text{min}}$ to denote 
 the coherent state evaluated at optimized Bloch angles, $\{\varphi_i^0\}_{i=1}^{2N_q}$.

The first derivative of the similarity transformed energy can be written as 
\begin{align}
  \label{eq:1st_der_formal}
  \frac{\mathrm{d}E(t)}{\mathrm{d} t}\Big|_{t = 0}
  & =
    \Braket{\frac{\mathrm{d}
    \boldsymbol
    \Omega_\text{min}}{\mathrm{d}
    t}| \hat H |\boldsymbol \Omega_\text{min}} +
    \Braket{\boldsymbol
    \Omega_\text{min}|
    \hat H |\frac{\mathrm{d} \boldsymbol
    \Omega_\text{min}}{\mathrm{d} t}} \nonumber \\
  & + \Braket{\boldsymbol \Omega_\text{min}|\frac{\partial \widetilde
    H}{\partial t}\Big|_{t=0}|\boldsymbol \Omega_\text{min}}.
\end{align}
Expanding the derivative of the mean-field state
$\mathrm{d} \ket{\boldsymbol \Omega_{\rm min}}/{\mathrm{d} t}$ as
\begin{equation}
  \label{eq:domega_dtau}
  \frac{\mathrm{d} \ket{\boldsymbol \Omega_{\rm min}}}{\mathrm{d} t} = \sum_{i=1}^{2N_q}
  \frac{\partial \ket{\boldsymbol \Omega_{\rm min}}}{\partial \varphi_i} \frac{
    \mathrm{d} \varphi_i}{\mathrm{d} t},  
\end{equation}
and accounting for the fact that all $\varphi_i$ are real, we can
write
\begin{align}
  \label{eq:1st_der_tr}
  & \Braket{\frac{\mathrm{d} \boldsymbol \Omega_\text{min}}{\mathrm{d} t}|
    \hat H|\boldsymbol \Omega_\text{min}} +
    \Braket{\boldsymbol \Omega_\text{min}| \hat H |\frac{\mathrm{d} \boldsymbol
    \Omega_\text{min}}{\mathrm{d} t}}  \nonumber \\
  & = \sum_{i=1}^{2N_q} \frac{
    \mathrm{d} \varphi_i}{\mathrm{d} t} \left[\Braket{\frac{\partial \boldsymbol \Omega_\text{min}}{\partial \varphi_i}|
    \hat H|\boldsymbol \Omega_\text{min}} +
    \Braket{\boldsymbol \Omega_\text{min}| \hat H |\frac{\partial \boldsymbol
    \Omega_\text{min}}{\partial \varphi_i}}\right] \nonumber \\
  & = 0.
\end{align}
The sum of derivatives in Eq.~\eqref{eq:1st_der_tr} vanishes because the expression in
square brackets is equal to 0 for any $i$ due to the condition of
$\varphi_i$ optimality. Thus, Eq.~\eqref{eq:PT1_test} follows from
Eq.~\eqref{eq:1st_der_formal} as
\begin{equation}
  \label{eq:partH_partt}
  \frac{\partial \widetilde H}{\partial t}\Big|_{t = 0} =
  -\frac{\I}{2} [\hat H, \hat P].
\end{equation}

To derive an expression for the second derivative, consider the
\gls{QCC} energy as a function of entangler's amplitudes and Bloch
angles [\emph{cf.} Eq.~\eqref{eq:barE}]
\begin{equation}
  \label{eq:eqcc_alt_notation}
  E_e(t, \boldsymbol \varphi) = \Braket{\boldsymbol \Omega | \widetilde H(t) | \boldsymbol \Omega}.
\end{equation}
Expanding $E_e$ into a multivariate Taylor series at $t = 0$ and
$\varphi_i = \varphi_i^0$ up to the second order gives
\begin{align}
  \label{eq:Ecc_2nd_full}
  E_e(t, \boldsymbol \varphi) & = E_\text{QMF} + t \frac{\partial
                                E_e}{\partial t} \Big|_{\substack{t = 0 \\
  \varphi_i = \varphi_i^0}} + \frac{t^2}{2} \frac{\partial^2
  E_e}{\partial t^2} \Big|_{\substack{t = 0 \\
  \varphi_i = \varphi_i^0}}
  \nonumber \\
                              &  + t \mathbf{c}^T\cdot \Delta{\boldsymbol
                                \varphi} + \frac{1}{2}\Delta{\boldsymbol
                                \varphi}^T \mathcal{D} \Delta{\boldsymbol
                                \varphi}, 
\end{align}
where
$\Delta{\boldsymbol \varphi} = \boldsymbol \varphi - \boldsymbol
\varphi^0$,
\begin{align}
  \label{eq:c_i}
  c_i & = \frac{\partial^2 E_e}{\partial t \partial
        \varphi_i}\Big|_{\substack{t = 0 \\   \varphi_i =
  \varphi_i^0}}, \\
  \label{eq:Dij}
  D_{ij} & = \frac{\partial^2 E_e}{\partial \varphi_i \partial
           \varphi_j}\Big|_{\substack{t = 0 \\ \varphi_i = \varphi_i^0}},
\end{align}
and the first partial derivatives
$\partial E_e/\partial \boldsymbol \varphi$ at $t = 0$ are zero
because we have chosen the origin of the Tailor expansion at the
mean-field solution. In fact, Bloch angles are minimized in
Eq.~\eqref{eq:ent_ranking_formula_appendix} for \emph{every} value of
the amplitude $t$, not only at $t=0$, which leads to the following
linear equations that determine the optimal set of Bloch angles for a
given $t$
\begin{equation}
  \label{eq:Bloch_minimum_cond}
  \mathbf{0} = \frac{\partial E_e}{\partial \boldsymbol \varphi} = t
  \mathbf{c} + \mathcal{D} {\Delta \boldsymbol \varphi}
\end{equation}
or
\begin{equation}
  \label{eq:Bloch_minimum_cond_2}
  {\Delta \boldsymbol \varphi} = -t\mathcal{D}^{-1} \mathbf{c}.
\end{equation}
Equation~\eqref{eq:Bloch_minimum_cond_2} can be interpreted as a linearized form 
of a nonlinear equation that determines minimizing Bloch angles as functions of the amplitude in
Eq.~\eqref{eq:ent_ranking_criterion}.

Inserting Eq.~\eqref{eq:Bloch_minimum_cond_2} into
Eq.~\eqref{eq:Ecc_2nd_full} and collecting terms of the same power in
$t$ gives the second-order derivative of $E(t)$ in terms of
derivatives in Eq.~\eqref{eq:Ecc_2nd_full}
\begin{equation}
  \label{eq:2nd_der_explicit}
  \frac{\mathrm{d}^2E_e(t)}{\mathrm{d}t^2}\Big|_{t =
    0} = \frac{\partial^2 E_e}{\partial t^2} \Big|_{\substack{t = 0 \\
      \varphi_i = \varphi_i^0}} - \mathbf{c}^T \mathcal{D}^{-1} \mathbf{c}.
\end{equation}
General expressions for the derivatives in the right-hand side of
Eq.~\eqref{eq:2nd_der_explicit} follow
\begin{align}
  \label{eq:d2f_dt2}
  \frac{\partial^2 E_e}{\partial t^2} \Big|_{\substack{t = 0 \\
  \varphi_i = \varphi_i^0}} & = \Braket{\boldsymbol
                              \Omega_\text{min} | \frac{\partial^2 \widetilde H}{\partial
                              t^2}\Big|_{t = 0} | \boldsymbol \Omega_\text{min}} \nonumber \\
                            & = \Braket{\boldsymbol
                              \Omega_\text{min}| \frac{1}{2}\hat P [\hat H, \hat P] |\boldsymbol
                              \Omega_\text{min}}
\end{align}
and
\begin{align}
  \label{eq:d2f_dt_dphi_i}
  c_i  & =  \frac{\partial^2 E_e}{\partial t \partial
           \varphi_i}\Big|_{\substack{t = 0 \\   \varphi_i =
  \varphi_i^0}} = \frac{\partial}{\partial \varphi_i}
      \left(\frac{\partial E_e}{\partial t}\right)  \nonumber \\
     & =  
      \Braket{\frac{\partial \boldsymbol \Omega_\text{min}}{\partial
      \varphi_i} | -\frac{\I}{2} [\hat H, \hat P] | \boldsymbol
      \Omega_\text{min}} \nonumber \\
         & \quad + \Braket{\boldsymbol \Omega_\text{min} |
           -\frac{\I}{2} [\hat H, \hat P] | \frac{\partial \boldsymbol
           \Omega_\text{min}}{\partial \varphi_i}},\\
  \label{eq:D_elem}
  D_{ij} =
  & \Braket{\frac{\partial^2 \boldsymbol
    \Omega_\text{min}}{\partial \varphi_i \partial \varphi_j} | \hat H |
    \boldsymbol \Omega_\text{min}} + \Braket{\frac{\partial \boldsymbol
    \Omega_\text{min}}{\partial \varphi_i} | \hat H | \frac{\partial \boldsymbol
    \Omega_\text{min}}{\partial \varphi_j} } \nonumber \\
  & + \Braket{\frac{\partial \boldsymbol
    \Omega_\text{min}}{\partial \varphi_j} | \hat H | \frac{\partial \boldsymbol
    \Omega_\text{min}}{\partial \varphi_i} } + \Braket{\boldsymbol
    \Omega_\text{min} | \hat H | \frac{\partial^2 \boldsymbol
    \Omega_\text{min}}{\partial \varphi_i \partial \varphi_j}}.
\end{align}

To evaluate the derivatives of the coherent states,
Eq.~\eqref{eq:spin_coh_state_qubit}, that appear in
Eqs.~\eqref{eq:d2f_dt_dphi_i} and \eqref{eq:D_elem}. Since the
coherent states constitute a full basis in the (two-dimensional)
Hilbert space of a single qubit, it is possible to express the
corresponding derivatives in the form of \emph{operators} acting on a
given coherent state. In particular, if $\theta_i$ is a polar angle of
the $i^{th}$ qubit, then direct evaluation of the corresponding
derivative from Eq.~\eqref{eq:spin_coh_state_qubit} gives
\begin{equation}
  \label{eq:dOmega_dtheta}
  \frac{\partial \ket{\Omega_i}}{\partial \theta_i} =
  \frac{1}{2}\left[-\sin\left(\frac{\theta_i}{2}\right)\ket{\alpha} +
    \E^{\I\phi_i} \cos\left(\frac{\theta_i}{2}\right)\ket{\beta} \right].
\end{equation}
This derivative can be also written as
\begin{align}
  \label{eq:dOmega_dtheta_op}
  \frac{\partial \ket{\Omega_i}}{\partial \theta_i}
  & = \frac{1}{2}\left[-\E^{-\I\phi_i}\hat s_{+}(i) + \E^{\I\phi_i}\hat
    s_{-}(i)\right] \ket{\Omega_i} \nonumber \\
  & = \frac{\I}{2}\left[\hat x_i \sin(\phi_i) - \hat y_i
    \cos(\phi_i)\right]\ket{\Omega_i}, 
\end{align}
where $\hat s_{+}(i) = (\hat x_i + \I \hat y_i)/2$,
$\hat s_{-}(i) = (\hat x_i - \I \hat y_i)/2$, $\hat x_i$, and
$\hat y_i$ are spin raising, lowering, Pauli $\hat x$, and $\hat y$
operators for the $i^{th}$ qubit, respectively. Analogously, the
derivative of $\ket{\Omega_i}$ with respect to the azimuthal angle
$\phi_i$ can be evaluated as
\begin{align}
  \label{eq:dOmega_dphi_op}
  \frac{\partial \ket{\Omega_i}}{\partial \phi_i}
  & = \I \sin\left(\frac{\theta_i}{2}\right) \E^{\I\phi_i}\ket{\beta}
    \nonumber \\
  & = \I \E^{\I\phi_i} \tan\left(\frac{\theta_i}{2}\right) \hat s_{-}(i) \ket{\Omega_i}.
\end{align}
Successive application of Eqs.~\eqref{eq:dOmega_dtheta_op} and
\eqref{eq:dOmega_dphi_op} gives access to higher-order derivatives of
the coherent states.

{\it Determining missing angles:} From \eq{eq:spin_coh_state_qubit} it is obvious that 
values of $\phi_i$ angles corresponding to $\theta_i=0$ or $\pi$ will not affect 
qubit mean field energy, and thus, the mean-field procedure does not define them. 
However, these angles will determine the values of the second derivatives
due to differentiation in \eq{eq:dOmega_dtheta_op}. To determine these $\phi_i$ 
angles we re-optimize all angles for small values of $\tau$ (e.g. $10^{-3}$) in the  
similarity transformed Hamiltonian. This procedure can be combined with a
derivative discontinuity test that we discuss next.

{\it Derivative discontinuities:} Another caveat of working with the derivatives 
of similarity transformed energies \eq{eq:ent_ranking_formula_appendix} is that 
occasionally, first derivatives have discontinuities in the 
$\tau=0$ point. In this case, the second derivative is not defined 
and should not be used. Such situations can be detected by evaluating 
QMF solutions for the similarity transformed Hamiltonian 
[\eq{eq:qcc_func_single_tau}] with small amplitudes (e.g. $\tau=\pm10^{-3}$)
and calculating the first derivatives in these points using \eq{eq:PT1_test}
with the $\hat H[\tau; \hat P]$ Hamiltonian instead of $\hat H$.
If the first derivatives are significantly different, this is a numerical indication 
of the derivative discontinuity, and the second derivative is not evaluated.

%

\end{document}